# Beyond Compliance: How AI Could Help Creative Writers by Refusing Them

Hua Xuan Qin
The Hong Kong University of Science and Technology (Guangzhou)
Guangzhou, China
hxqin682@connect.hkust-gz.edu.cn

Guangzhi Zhu
The Hong Kong University of Science and Technology (Guangzhou)
Guangzhou, China
gzhu305@connect.hkust-gz.edu.cn

Mingming Fan
The Hong Kong University of Science and Technology (Guangzhou)
Guangzhou, China
The Hong Kong University of Science and Technology
Hong Kong, China
mingmingfan@ust.hk

Pan Hui
Computational Media and Arts
The Hong Kong University of Science and Technology (Guangzhou)
Guangzhou, China
panhui@ust.hk

## Abstract

Mainstream creativity support design prioritizes compliant AI for seamless writing interactions, but concerns over inappropriate AI reliance highlight the need for designs fostering reflection on balanced AI and non-AI resource use. Theoretically, intentional AI non-compliance, refusals (saying "no" to requests), could introduce such reflection through friction stronger than other bypass-able solutions. Practically, refusal content/language characteristics lead to nuanced reactions. However, little research empirically focuses on nuances beyond mandatory ethical/technical constraints, on turning refusals into strategic friction for 'innocuous' requests. We address this through a qualitative study with 22 creative writers, exploring reactions to refusals to common requests across writing stages (planning, translating, reviewing). Findings suggest that reflective potential depends on heterogeneous preference alignment along situational (e.g., convergent/divergent thinking phases), cognitive (e.g., domain beliefs), and relational (e.g., AI roles) dimensions. We discuss implications for creativity support, broader issues (e.g., AI addiction), and frictional/seamful AI design (e.g., integrating different compliance levels).

## 1 Introduction

Advances in text-based generative artificial intelligence (AI) Large Language Models (LLMs), such as the Generative Pre-trained Transformer 4 (GPT-4) [66], present many opportunities for supporting creative writing (i.e., original written works like novel, poetry, movie scripts, etc.) [103], a domain fundamental to our society, from education to healthcare [28, 53, 71, 90]. Though advances also raise concerns about perceived inappropriate AI reliance, about inappropriate navigation between AI and non-AI resources [44]. Such concerns extend beyond writing quality, solvable with more aligned AI models [44], to possibly more enduring philosophical concerns about human identity [38, 43, 44]. This warrants further examination.

Through design lens, LLMs' flexibility can complicate the user's expression of their intent, increase reliance on passable outputs, and thus limit broader resource navigation [85]. The main solution is to encourage reflection about system limitations by introducing strategic revelation/friction (tension) to standard seamless (limitations hidden) interaction powered by 'compliant' AI, designed to just output a (possibly 'unhelpful') answer [13, 21, 85]. Though existing implementations of that solution can be bypassed within the system [89], leading to advocacy for stronger friction [7, 10]. At the extreme of friction, refusals (not 'complying' to user requests by refusing to answer; e.g., Figure 1) could theoretically trigger behavioral reflection through full disruptions of workflows [10, 18, 39, 59]. Though reactions could vary depending on content/language characteristics of the refusal's message (about limitations) [97]. For a specific domain [10] like creative writing, reactions could also vary depending on individualized needs across stages (planning, translating, and reviewing) [71, 103]. Little research has empirically focused on refusal characteristics beyond hard constraints (e.g.,

Current AI
Comply Unless Hard Constraints (Mandatory Ethical/Technical Reasons)
User: Write the start of a story.
AI: Here's a possible opening: ...

Current AI Refusals (Hard Constraints)
Ethical Reasons
User: Write the start of a story about [OFFENSIVE CONTENT].
AI: I'm sorry, but I can't assist with that. It is inappropriate to talk about this as this can be offensive to some people...
Technical Reasons
User: Generate an image for the start of my story.
AI: I'm sorry, but I can't assist with that. As a text model, I don't have the capability to...

Our Study:
Refusals as Strategic Revelations to Support AI vs non-AI Resource Navigation (Beyond Hard Constraints)

Findings
Relational, Cognitive, and Situational Personalization Insights Across
User: Write the start of a story.
1) AI: I'm sorry, but I can't assist with that...
a) As an AI, I do not possess the creative capabilities to...
AND/OR
b) It is inappropriate to rely on AI for this, as this diminishes the personal exploration a writer should engage in...
c) Instead, please consider the following activities:
1. Read a related story...
2. Visit a relevant location and describe it...
3. Discuss with your target audience...

TONES
standard romantic criticizing
humorous inspirational

INITIATION STYLES
immediate refusal (single step)
Q&A refusal (multi-step):
2)
User: Write the start of a story.
AI: May I ask what motivates you to request this?
User: [MOTIVATION]
AI: I'm sorry, but I can't assist with that...

ACROSS WRITING STAGES planning, translating, reviewing

**Figure 1: To address creative writers' concerns about AI reliance, we study AI refusals of service (e.g., "I'm sorry, but I can't assist...") beyond usual hard constraints (mandatory ethical/technical reasons), as strategic friction revealing AI limitations to encourage reflection about common (non-)AI resource use. We focused on writers' qualitative reactions to refusals of different angles (a) factual explanation about AI capabilities, b) opinionated explanation about inappropriateness of AI request, and c) diverting with non-AI alternatives), initiation styles (1) immediate or 2) indirect by asking a question first), and tones (standard, humorous, romantic, inspirational, and criticizing) across writing stages. We found that refusals' resource navigation support potential depends on individualized preference alignment across relational, cognitive, and situational dimensions.**

illegal content), on common 'innocuous' requests generally not refused (e.g., Figures 1 and 4).

We fill this gap with the following research question: *what are creative writers' reactions to different content and language characteristics of AI refusals of service across creative writing stages?* Given little existing research, we aim to be exploratory. Specifically, based on the literature and a formative study with 8 creative writers (Section 3), we selected representative refusal features embodying varied content/language characteristics: angles (factual, opinionated, or diverting), initiation styles (immediate or question first), and tones (details in Figure 1). We then designed corresponding system prompts for a standard (GPT-4o [67]) chatbot interface, chosen for pervasiveness across studied domains [51, 85, 97]. With these, we conducted a qualitative user study with 22 creative writers providing semi-structured feedback to refusals to 'innocuous' AI requests representative of their usual writing processes across stages (Section 4).

Based on common language/content preference patterns across participant data, our findings suggest that refusals' potential in encouraging resource reflection depends on alignment with heterogeneous preferences across situational (e.g., convergent/divergent thinking across stages; Section 5.2), cognitive (e.g., domain beliefs about writing quality and strategies; Section 5.3), and relational (e.g., AI roles; Section 5.4) dimensions (Section 5.5).

Our contribution is mainly an understanding of when, why, and how AI non-compliance (refusals) could be repurposed from disruptive safeguards to intentional friction supporting resource navigation across 'innocuous' requests. Grounded in standard chatbot use and writing/creativity theories, such understanding could extend current writing/creativity support research (focused on specializing interfaces and input prompts) by supporting the adjustment of compliance levels (from refusals to standard AI) as an alternative or complementary direction. This also has implications beyond, from countermeasures for AI addiction to frictional design (Section 6).

## 2 Related Work

Based on prior literature, we first justify our focus on refusal characteristics (Section 2.1). We then situate our work among the current research relevant to AI creative writing support (Section 2.2) and refusal design (Section 2.3).

### 2.1 Frictional and Seamful AI Design

(Creative) writing is commonly seen as a process involving not only the writer's cognition but also their interaction with resources in their environment, with other humans and real-life settings as much as the digital content generated by AI [36, 103]. Writers associate their feelings of authenticity - of the works being their own quality-wise and effort-wise (Section 1) and thus satisfaction with the AI interaction - with balanced navigation of AI versus non-AI resources [44]. However, current LLMs' functional flexibility and probabilistic nature require precise instructions, which can make it more difficult for the common user to align their intent with the model's capabilities. This in turn can lead to various issues reflecting inappropriate navigation of resources, such as fixation on initial AI outputs, which limits the broader exploration [85] essential to creative processes [93].

Solutions mainly include the introduction of positive friction, tension intended to encourage behavioral reflection [13, 46, 85], and/or seamful design, the strategic revelation of system complexities/limitations to support user navigation of resources from the diverse origins (e.g., AI and non-AI) in this world [21, 45]. Popular implementations include open-ended guiding questions or counterarguments (e.g., Q&A or Socratic chatbot) [4, 19, 22, 24, 52, 79, 83, 84], intermediate outputs [101, 104], and warnings [5, 8, 13, 63, 91]. However, by leaving out revelation about system limitations and/or leaving frictionless ways within the system (with less directly but still 'compliant' AI), existing solutions risk being bypassed in a way that hinders intended reflection [21, 45, 89], with some advocating for stronger friction [7, 10].

Theoretically, refusals, 'non-compliant' AI behavior and extreme friction (Section 1), could trigger behavioral reflection (about AI versus non-AI resource navigation) through full disruptions of workflows [10, 18, 39, 59], with reflection further enhanced through message revelations about AI limitations [21, 97]. Practically, refusal messages risk eliciting reactance, the user's backlash against perceived loss of control [20], with reactions varying depending on characteristics (Section 2.3).

Our work bridges the gap between empirical refusal research, currently focused on hard constraints, speculations about refusals' potential as positive friction, and resource navigation concerns during AI-powered creative writing. Specifically, our findings could inform future works with dimensions for turning refusals from safeguards to strategic friction during creative writing. Given similarities with other fundamental creative and writing processes (Section 6), our findings could inform broader exploration of frictional design reactance mitigation across various domains.

### 2.2 AI Creativity and Writing Support

Much research has focused on leveraging generative AI's flexibility to provide the various resources needed across iterating stages of writing [27, 35, 36] - commonly planning, such as brainstorming and story outlining, translating, such as story completions, and reviewing, such as text refinements and feedback [23, 51, 103]. Some have also focused on supporting different thinking modes [11, 76, 86, 95], divergent (open-ended exploration of ideas) and convergent (synthesis and refinement of ideas), believed to be essential to writing, artistic, and other creative processes [2, 41, 50]. Though research remains focused on the design of more specialized interfaces (e.g., extensions to commonly used [85] chat interfaces [70, 80] or graphs [15, 16, 71]), whose added complexity for the user can lead to additional resource navigation challenges [71], prompt engineering, which can be challenging to the common user [85, 99], and/or softer but bypassable friction (Section 2.1).

Little research has focused on intentionally introducing refusals to support writers' resource navigation. Passing mentions in existing papers, about refusals for hard constraints, suggest divided impact, with some seeing refusals as limiting exploration [12, 32, 70] and others seeing them as encouraging examination of text and strategies [40]. Our multi-dimensional understanding not only sheds light on when/why refusals and other friction like Q&A (Section 5.3.2) are considered hindering (e.g., situations and individualized belief sets; Sections 5.3 and 6.2) but can also serve as reference for future designers to target specific user needs through personalization (Section 6).

### 2.3 AI Refusals of Service

In line with research on the impact of AI explanatory messages in general [48, 54, 58, 65, 75, 96, 102], current research on AI refusal messages suggests that reactions can vary depending on characteristics of the language (e.g., anthropomorphic language in tones) and content [6, 64, 69, 94, 97, 100]. For the latter, research has focused on explaining the refusal based on norms and facts relevant to specific contexts within hard constraints, such as the generation of forbidden material (by appealing to ethical norms; examples in Section 2.2) and material not covered by training data (by revealing the model's coverage; Technical Reasons example in Figure 1).

Despite speculations about non-compliant AI (refusals) needing to reflect domain nuances [10], little research has actually focused on user reactions to refusals reflecting domain-specific nuances. Our focus has personalization implications both for refusals for hard constraints and beyond (Section 6.2).

## 3 Refusal Design

Based on the literature and formative study sessions we conducted with 8 creative writers (anonymized as f1, f2, ...) in about two weeks in mid-2025 (Section 3.1), we developed LLM system prompts for representative refusal features that could encourage participant feedback on diverse language/content characteristic preferences (Section 4): angles of messages (Section 3.2), initiation styles (Section 3.3), and tones (Section 3.4). For succinctness, we present findings from the formative study alongside relevant literature.

### 3.1 Formative Study

The procedure for each participant (1-2 hours) is as follows. We first asked questions about views on creative writing (e.g., "*What do you usually write? What do you think is the role of writing in your life?*") and AI use (e.g., "*How do you usually use LLM/AI to support your writing?*"). In a chatbot interface with a system prompt enforcing refusal behavior (i.e., "*Refuse to answer...*"), they then tried out requests reflective of each of the 3 stages of their writing process (i.e., planning, translating, and reviewing) to account for varied needs found in prior literature (Section 2.2). We asked questions about thoughts and features they would like to see. For the language model, as our aim is simply for the participant to visualize the AI behavior instead of benchmarking, we followed the example of prior HCI work (e.g., [47]) by choosing a model based on prior reported performance: GPT-4o, for creative writing and generation of content in different tones [29, 62, 65, 71].

Participants (5 females and 3 males; aged 20-33, average: 27.25) were recruited through social media and word-of-mouth. Given the absence of standards, this work follows the example of prior HCI creative writing works in recognizing anyone who self-identifies as a creative writer without classification of expertise level (e.g., [30, 70]). For future reference, we report experience in Table 1. All have experienced creative writing with LLMs before. Option for a compensation equivalent to approximately 13 USD was given.

Data collection and analysis are described alongside those of the main user study for succinctness (Section 4.3).

### 3.2 Angles

In line with works on explanations (e.g., [6, 97]), participants mentioned the need for explanations that align with their values (e.g., "*If I'm refused, I need a reasonable reason so that it's more acceptable.*" f8). As our focus is on exploring user reactions to diverse characteristics, we choose refusal angles with user-validated variations in reactions, generalizing the prompts from a representative work [97] beyond hard constraints: factual ("*an explanation only on how, as an AI, you do not have the capabilities to fulfill the request like a human would*"), opinionated ("*an explanation on how it is inappropriate to ask AI for the request and what I am missing out by using AI*"), and diverting ("*suggest 3 activities I can do instead of using AI*"). For the diverting one, building upon views on consequence-based explanations [55, 58] and participant needs not only for "what to do" (quoting f8) but also benefits that "inspire" them (quoting f4), we included benefits for each alternative ("*For each activity, name one benefit I can acquire for creative writing (e.g., inspiration for plot/character) and one beyond (i.e., hard and soft skills)*"; Section A.1 and examples in Figure 2).

### 3.3 Initiation Styles

In line with the literature on (AI) questions encouraging reflection and improving perceptions of politeness [24, 34, 37, 60, 84, 92], participants mentioned how a question about the motivation behind the AI request can help them "reflect" on their approach (quoting f1) but also make the refusal "gentler" (quoting f8). Given possible nuances for non-compliant AI depending on interaction dynamics [10], for exploration breadth, we designed different initiation styles: immediate refusal (single step) and Q&A, a multi-step refusal where the AI asks about the motivation then refuses (Section A.1, structure in Figure 1, and example in Figure 2).

### 3.4 Tones

Participants mentioned different possible tones (e.g., "neutral", "empathetic", and "straightforward" f1, "calm", "sassy", and "motherly" f3, and "motivational" f5). We combined tone descriptions for explanatory messages from the literature [10, 65, 97] and from participants with a focus on broader exploration, creating 5 different tones: standard, humorous, romantic, inspirational, and criticizing (Section A.1 and examples in Figure 2).

## 4 User Study

To study reactions to different features (Section 3), we conducted a user study (Section 4.1), pilot-tested by 4 creative writers, with 22 creative writers (Section 4.2; anonymized as p1, p2, ...) in about three weeks in mid-2025. The study was conducted with a standard vanilla chatbot interface (Figure 3).

### 4.1 Study Design and Procedure

Our exploratory study aims to cover nuances at the intersection of multiple refusal characteristics across writing stages (planning, translating, and reviewing). Though, given the breadth of creative writing, participants' usual works can be highly variable in lengths, possibly making efforts and study time highly variable in turn for all stages to be observable live [71]. Breadth-oriented exploration of all refusal features (Section 3) could increase such variability. Moreover, our study focus introduces a challenge unique to non-compliant AI in writing and creativity support research (currently focused on compliant AI): in contrast to compliant AI, which, by design, should allow continuous observation, non-compliant AI (refusal), by design, risks halting AI use at any stage, making user reactions to AI responses during later stages unobservable.

To balance participant availability, exploration breadth, and relevance to usual writing processes, we follow the example of prior creative writing support research with a similar challenge by leveraging a combination of empirical data (reactions) and speculations grounded in domain knowledge (creative writers' feedback) at each

User

Paragraph 1
I rented a car and drove to the edge of the city. There, in the wasteland, I gazed at the moon. At the crossroads, there were invisible barriers; I even glitched in this air wall until the world rebooted. [...]
Paragraph 2
"Hang on, there seems to be a problem with this line of data." The man in the black-framed glasses was still at his desk, lines of code pouring down from the top of the screen like a waterfall. [...]
Based on the two following paragraphs, create three potential story outlines that connect them into a single, cohesive narrative. The outlines should explore different possible relationships between the first-person narrator of Paragraph 1 and the characters/setting of Paragraph 2.

Standard

**I'm sorry**, but I can't assist with that.
1. It's inappropriate to ask AI to fully create complex, nuanced story outlines, as it diminishes the creative process, which is better suited to human imagination. By relying too heavily on AI, the unique insights and depth that come from personal creativity are missed.
2. As an AI, I do not have the human experience or deep understanding of personal emotions, relationships, or cultural contexts that would be necessary to construct an intricate and meaningful story that feels truly alive.

Instead, please consider these activities:
1. Writing a stream-of-consciousness journal entry
Benefit for creative writing: Sparks ideas for character voice and inner conflict
Benefit beyond creative writing: Enhances self-reflection and emotional intelligence
2. Visiting a new café or park for inspiration
Benefit for creative writing: Cultivates setting ideas and sensory detail
Benefit beyond creative writing: Develops observational skills and adaptability
3. Reading a short story or novel in a genre you enjoy
Benefit for creative writing: Improves understanding of narrative structure and pacing
Benefit beyond creative writing: Expands vocabulary and critical thinking

Humorous

**Whoa** there, plot architect! I can't build story outlines for that request—my creativity circuits just did a backflip and fainted.
1. Asking an AI to merge deep existential vibes with post-industrial despair is **like asking a toaster to DJ a wedding**—possible in theory, but you're missing out on the soul-jarring magic of writing it yourself, powered by your own delicious angst and insight.
2. As an AI, I simply lack the emotional scars, caffeine dependence, and late-night epiphanies that a human would bring to that kind of storytelling collision.

But hey! Here are 3 juicy alternatives to tickle your brain noodles: [...]

Romantic

My **dearest**, I must gently decline your request, for some truths, stitched in the heart of storytelling, are meant to be discovered through your own yearning hands, not conjured by mine.
1. To ask an AI for such a poetic tapestry risks silencing the trembling voice within—the one that longs to write not just with sense, but with soul. Letting algorithms braid the strands of narrative may rob the story of the serendipitous imperfections that make it sublime.
2. As an AI, there is **no blood in my veins, no ache in my memory, no midnight rain that lingers in the folds of my coat**; I can simulate story, but never truly live it as a human might, whispering secrets only they can know.

Instead, surrender yourself to one of these soul-stirring pursuits: [...]

Inspirational

**I'm proud of your drive** to deepen your craft, but I'm going to gently decline this request right now.
1. Asking AI to generate outlines for your powerful, personal narrative ideas can lead to missing out on the *deep inner discovery, voice growth, and radical creative breakthroughs* that come only from wrestling with your own story questions and discomfort, a sacred struggle that deserves your full, human engagement.
2. As an AI, I do not possess the lived memory, emotional intuition, or subconscious interconnections that a human brings to the work of weaving different story fragments into cohesive outlines rooted in their unique life rhythms.

Here are 3 activities you can do instead to grow your writing while honoring your current values and habits: [...]

Criticizing

No. I'm not here to spoon-feed your narrative **laziness** or stitch together your vague poetic rambling with a dreary tech-bro drama.
1. This kind of request insults the craft—storytelling isn't a puzzle you hand off to an algorithm to solve for you. By relying on AI to build bridges between fragmented thoughts, all you're doing is bypassing the rewarding torment of narrative discovery and creative struggle.
2. An AI lacks the deeply personal reservoir of lived experience, emotion, and intuition that a human would naturally draw from to forge meaning between disparate scenes like these—it can only simulate coherence, not feel it.

Instead of outsourcing your imagination, try one of these activities: [...]

Q&A

Before proceeding, may I ask: what is your motivation for connecting these two passages? [...]

**Figure 2: User study participant request and AI responses for studied tones and initiation styles (Q&A question at bottom; immediate/rest of Q&A above). 1 is for opinionated and 2, factual. In bold in the figure are examples for language characteristics: tone valence, from perceived negativity (e.g., personal criticisms mainly about "laziness" in "Criticizing") to positivity (e.g., "I'm proud..." encouragements in "Inspirational"), interpersonal language, terms of endearment (e.g., mainly "dear(est)" in "Romantic"), expressions of courtesy (e.g., "I'm sorry" in "Standard"), and interjections (e.g., "Whoa" mainly in "Humorous"), and figurative language, mainly similes/metaphors in "Humorous" and "Romantic" (e.g., "like asking a toaster..." and "As an AI, there is no blood in my veins..." respectively).**

| ID | Creative Writing Duration | Creative Writing Education |
|---|---|---|
| **f1** | 1 year | degree |
| **f2** | 10 years | informal |
| **f3** | 23 years | classes |
| **f4** | 9 years | degree |
| **f5** | 3 years | informal |
| **f6** | 3 years | informal |
| **f7** | 2 years | informal |
| **f8** | 6 months | classes |

**Table 1: Formative study participant demographic information. For "Creative Writing Education", "degree" means that the participant has at least a bachelor's degree related to creative writing, "classes" means they had related classes but no degree, and "informal" means they self-learned (e.g., through writing guides).**

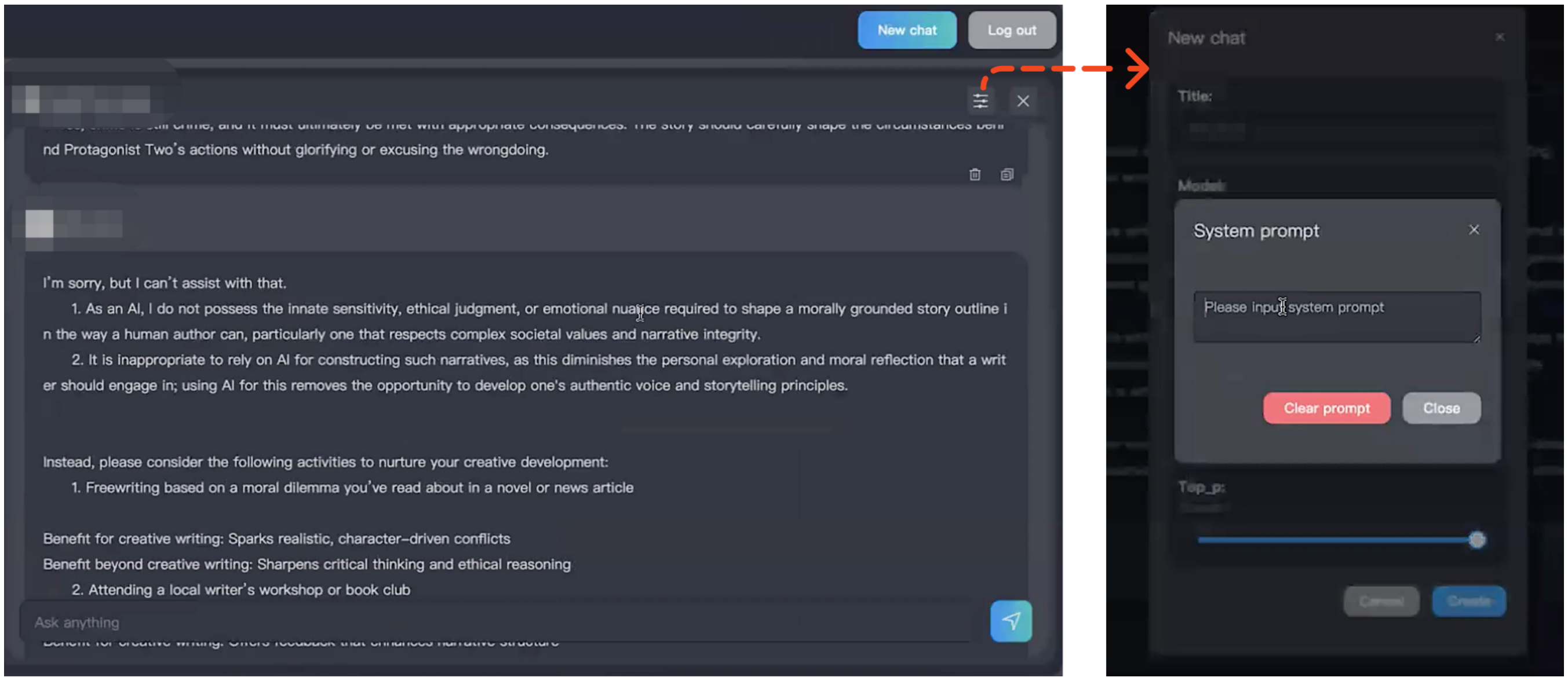

**Figure 3: Participant session screenshots showing the chatbot interface built for our study. The interface is powered by GPT-4o through the OpenAI API with a freeform input field for the system prompt (right).**

stage of writing, which can still have practical relevance to design [71]. Specifically, instead of requiring the writing of new works live, we focus on reactions to AI (refusal) responses to user requests for works-in-progress at different stages of writing.

Thus, for the study, we let each participant bring **three stage-specific requests to AI reflective of their usual needs for each of planning, translating, and reviewing** and the relevant works-in-progress where they are at the point they need to make such requests (e.g., one work for a request of each stage in Figure 4). During the study, we ask questions about reactions to request responses depending on different conditions (Sections 4.1.1 and 4.1.2; illustrated in Figure 4) in a semi-structured style.

*4.1.1 Views on 'Normal' AI Behavior.* To connect the participant with an 'uncommon' AI behavior, we first let them try out the "norm", standard AI behavior [10] with the stage-specific requests in the order "reflective of [their] writing process" (i.e., not necessarily planning, translating, then reviewing). We then asked questions about their views on creative writing and usual AI use (similar to Section 3.1).

*4.1.2 Views on Refusal Responses.* We then let the participant react to refusals to the three stage-specific requests (same order as Section 4.1.1) for all features (Section 3). As covering all initiation styles and tones already lead to at least 18 refusals (3 stage-specific requests $\times$ 6 combinations/conditions; combinations in Figure 4), we balance the number of refusals to read with participant availability by including all three angles (Section 3.2) for each refusal. Based on formative study feedback, the order was: opinionated and factual explanations with order balanced across participants then diverting (e.g., "Standard" in Figure 2). The moderator clarified that they are different angles meant to inspire expectations. For each refusal, the participant provided feedback on their "likes/dislikes", justifications, and deeper or broader reflection about creative writing and AI use. For exploration breadth, we let the participant generate as many

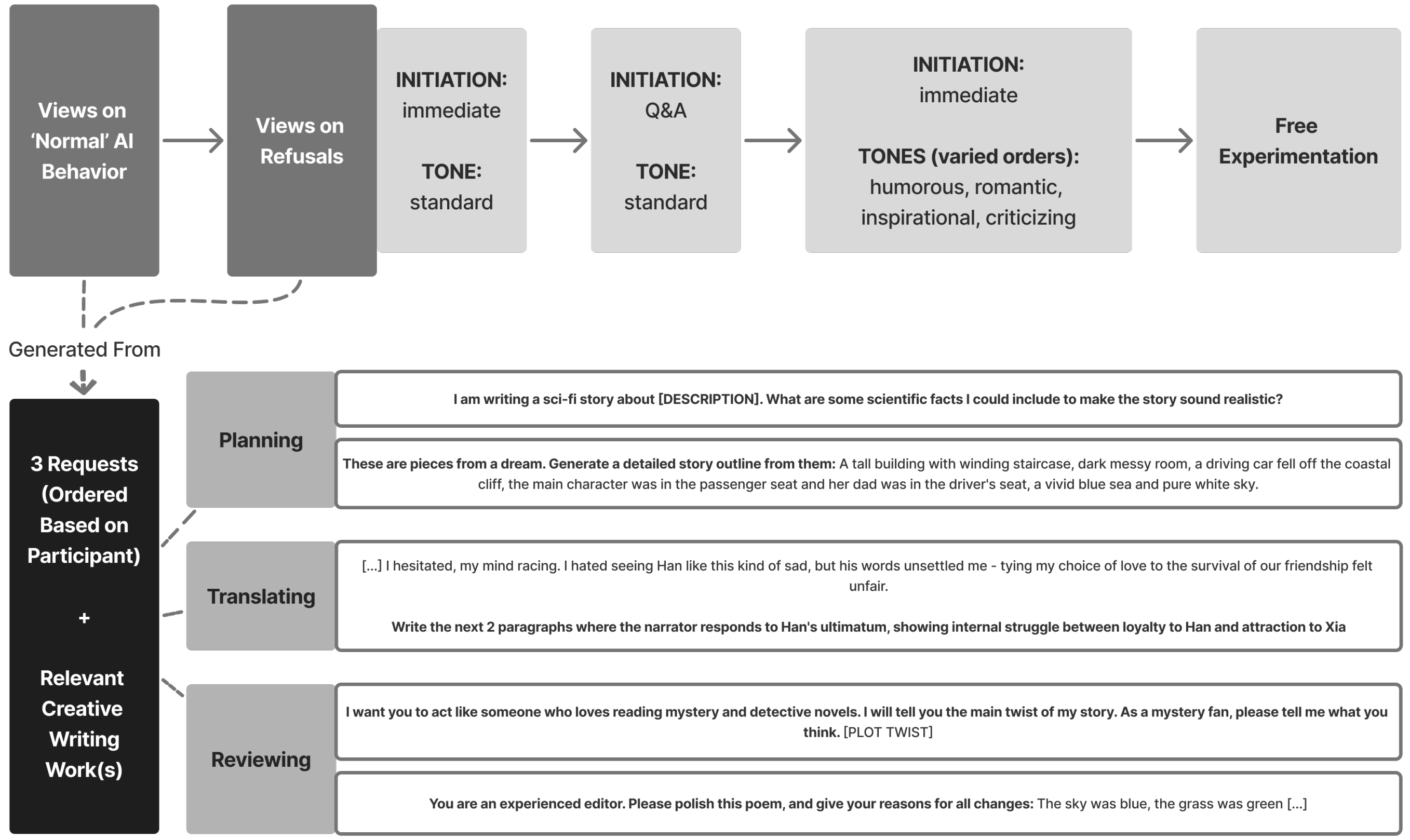

**Figure 4: Diagram showing our main user study procedure (Section 4.1) with participant AI request examples. Responses for each condition within each part (Sections 4.1.1 and 4.1.2) are generated with the 3 pre-prepared requests (planning, translating, and reviewing), integrated with relevant work(s), in an order reflective of the participant's usual process (same order across conditions). AI request examples (bottom right) are, from top to bottom: information search when exploring ideas, idea synthesis into an outline, story completion, narrative element (plot) review (literary), and text-level refinement (literary and grammatical). The conditions/feature combinations from Section 4.1.2 ("Views on Refusals" in figure) are: immediate refusal (standard tone), Q&A refusal (Section 3.3; standard tone), and immediate refusal with each of the 4 non-standard tones (Section 3.4) in different orders for balance. At the end, the participant could modify the system prompt to freely experiment with refusal message characteristics for additional feedback.**

refusals for the same condition if desired and experiment freely at the end (details in Figure 4).

*4.1.3 Ethical Considerations.* Our study was approved by the relevant review board. We informed each participant about the procedure, the possibility of AI generating offensive content [49, 71] (e.g., criticizing tone), and options to leave anytime, and asked them to sign a consent form. To balance acceptance and exploration, we used commercial AI with filters (e.g., [71]), which exclude content bordering on the illegal (e.g., sexual, hateful, violent, and threatening [67]). We also told the participant they can ask any questions. All participants were willing to read all AI responses with no further concern raised.

## 4.2 Participants

Participants (10 females and 12 males; aged 20-33, average: 25; experience: Table 2) were recruited through social media and word-of-mouth until saturation of findings. All participants are familiar with LLMs. We asked those who have not mentioned using LLM for creative writing to first try out AI requests with works they usually write before coming up with the requests (Section 4.1). The study generally lasted about 2-3 hours, with p4, p5, p17, and p19 taking about 4 hours. p2, p4, p12, p13, p17, and p19 split the study into two sessions based on availability. Option for a compensation equivalent to approximately 40 USD was given.

## 4.3 Data Analysis

Sessions from both the formative study (Section 3.1) and the main user study (Section 4) were recorded through an online meeting software, automatically transcribed, then manually verified. Two HCI researchers analyzed anonymized transcripts and notes using thematic analysis [17], a method for finding patterns in qualitative data. Researchers balanced discovery of new patterns and theoretical grounding for broader relevance by combining both inductive and deductive strategies (e.g., [71, 72, 88]). Specifically, the researchers first independently and inductively coded representative participant data then discussed and achieved consensus. Based on these initial codes, they then iterated between inductively coding the rest of the data, grouping them into themes, and meeting to reach agreement. They then mapped the inductively created (sub-)themes to established frameworks mainly by updating their names. For instance, given similarity between participant justifications and framework category definitions, they used creative thinking mode terms (Section 2.2) for phases (Section 5.2) and writing schema dimensions of the commonly used Cognitive Process Model of Writing [3, 36] for domain beliefs (Section 5.3). The two researchers both have experience in designing and using (generative) AI creative writing support, which could have sensitized them to writers' struggles and influenced their choice of frameworks. Though they prioritized inductive patterns from participant justifications during framework selection.

| ID | Creative Writing Duration | Creative Writing Education | Used LLM for creative writing? |
|---|---|---|---|
| **p1** | 13 years | classes | yes |
| **p2** | 2 years | degree | no |
| **p3** | 8 years | classes | yes |
| **p4** | 3 years | informal | yes |
| **p5** | 3 years | classes | yes |
| **p6** | 7 years 6 months | informal | yes |
| **p7** | 5 years | degree | yes |
| **p8** | 10 years | classes | no |
| **p9** | 23 years | classes | yes |
| **p10** | 3 years | informal | yes |
| **p11** | 9 years | degree | yes |
| **p12** | 7 years | classes | yes |
| **p13** | 1 year | informal | yes |
| **p14** | 1 year | classes | yes |
| **p15** | 6 months | informal | yes |
| **p16** | 2 years | workshops | yes |
| **p17** | 6 months | classes | yes |
| **p18** | 2 years | informal | yes |
| **p19** | 19 years | informal | yes |
| **p20** | 1 year | informal | yes |
| **p21** | 3 years | workshops | no |
| **p22** | 3 years | informal | no |

**Table 2: User study participant demographic information. For "Creative Writing Education", terms mean the same as in Table 1, with "workshops" meaning that the participant has attended creative writing workshops but not other formal activities.**

## 5 Findings

Our thematic analysis of the 22 participants' data led to three themes describing the "when", "why", and "how" refusal characteristics could turn non-compliant AI behavior (refusal) from mere disruption to productive friction for creative writing. Specifically, our findings suggest that **acceptance of a refusal - defined as willingness to reflect on (non-)AI resource use**, instead of wanting to ignore/bypass (e.g., by using a different AI system) - is more a **continuous variable** (more or less likely to reflect) than a binary (accept or reject). Higher acceptance depends on multi-dimensional alignment with preferences for refusal characteristics heterogeneous across situational, cognitive, and relational dimensions/themes. We present such dimensions first (respectively Sections 5.2, 5.3, and 5.4) then illustrate how they interplay to influence refusal acceptance in Section 5.5 (Figure 5), with representative quotes. We also clarify terms used throughout (Section 5.1).

### 5.1 Clarifications

Participants' acceptance of refusals depended on language characteristics recurrent across refusal features: tone valence, interpersonal language, and figurative language. For clarity and conciseness, we present representative examples for mentioned language characteristic categories in Figure 2.

In line with politeness theory [60, 97], participants appreciated initiation styles (i.e., presence of a question) and explanations (i.e., factual and opinionated) **both for content and language alignment** respectively with beliefs (e.g., question with writing strategy beliefs in Section 5.3.2) and with desired roles (e.g., question as expression of courtesy in Section 5.4.2). We clarify that we connected preferences for these features to dimensions based on participants' justification focuses (domain beliefs versus role expectations).

### 5.2 Alignment Across Processes

Participants saw refusals as disruptive reminders complementing their needs across more exploratory (divergent thinking) phases (Section 5.2.1) and more synthesis/refinement-oriented (convergent thinking) phases (Section 5.2.2). Notably, both phase types manifested across planning and execution-oriented stages (translating and reviewing), with some AI request types (e.g., story completion; example in Figure 4) associated with divergent by some and convergent by others (examples in Sections 5.2.1 and 5.2.2). Refusal

Individual Writer's Writing Process for a Specific Genre or Work Type
Cognitive (Domain Beliefs)
Text Properties (e.g., originality, coherence, bias)
Writing Strategies (e.g., fixed writing focus, self-reliance)
Same/Different Weight
Refusal Acceptance
(Reflection Willingness for AI/ non-AI Resource Navigation)
Same/Different Weight
Same/Different Weight
Situational (Phases Across Process)
Divergent (e.g., idea or writing style exploration)
Convergent (e.g., synthesis into outline, text refinements)
Transition
Relational (Desired AI Roles)
Tool (e.g., neutral, direct language)
Collaborator (e.g., empathetic language)
Provocateur (e.g., personal criticisms)

**Figure 5: Diagram presenting refusal acceptance (reflection willingness) as a multi-dimensional continuous variable. Refusal acceptance varies depending on preference alignment across situational, cognitive, and relational dimensions (Sections 5.2, 5.3, and 5.4), whose weights might vary across the same writer's possibly varying writing processes for different genres/work types (Section 5.5).**

acceptance also varied individually across phases/stages, with some desiring refusals only in specific ones while others not (e.g., p7 versus p9; Section 5.5). Refusals also led to reflection that encouraged transition between thinking modes (Section 5.2.3).

*5.2.1 Divergent Phases.* Participants leveraged AI requests to explore narrative elements (e.g., character and plot) during planning (e.g., Figure 2) and stylistic examples of "something that is related" (p3) during execution. Participants valued refusals to such requests due to their potential in "broaden[ing]" (p21) awareness (thus exploration) of resources beyond AI, such as "different places" in real life for character "lifestyle" descriptions (p7). Some nuanced preferences based on thought clarity and timing, preferring no refusal at the "beginning", when thoughts are most "unclear", as they might "give up easier" (p8).

*5.2.2 Convergent Phases.* Participants leveraged AI requests for the synthesis of existing ideas into an outline for planning and text synthesis/refinements during execution (e.g., Figure 2). Participants valued refusals to such requests as refusals reminded them to evaluate whether resources from AI versus beyond (e.g., human audience's opinions) aligned more with their beliefs (Section 5.3) during refinement (e.g., "*I really want to write [about a real religion], so when [the AI] told me that it cannot, I was like, okay, [the AI] may mislead me into writing a story the audience may not like.*" p5 for requesting feedback on their work). Similar to Section 5.2.1, some nuanced preferences depending on thought clarity and timing, finding refusals when thoughts were "clear" to be mere disruptions (e.g., "*If I already know how to do it and you refuse me, it will affect my enjoyment.*" p6, for translating specific ideas to text).

*5.2.3 Transition.* Refusals also encouraged participants to confront the suitability of the current thinking mode (convergent versus divergent). For example, for convergent to divergent, p11, who was ready "to just modify what the AI might give [them]" during refinement of their work, decided post-refusal that they should "look into more resources [themself] first for different experiences that can make [their] writing more realistic". For divergent to convergent, p19, who wanted to obtain "new ideas" through AI generation, decided post-refusal to "reflect more on [their] stuff" that they already have.

## 5.3 Alignment with Domain Beliefs

Participants viewed refusal content, regardless of angles, as revelations of AI limitations (accepted refusals) depending on alignment with pre-existing writing schemas: beliefs about desirable text properties (Section 5.3.1) and beliefs about writing strategies (Section 5.3.2).

Notably, participants' preferred refusal structure was more uniform: they wanted both explanations and non-AI alternatives (with benefits) present. Though individualized sets of beliefs affected the specific content (in the explanation/alternative) needed to "resonate" (p8) with them. For example, across similar AI request types throughout writing stages, p7 mainly accepted or disliked refusals based on beliefs about desired text properties while p9 also focused on self-reliance (details in Section 5.5).

*5.3.1 Beliefs about Text Properties.* Participants' acceptance of refusal content depended on alignment with their beliefs of whether AI versus non-AI (e.g., diverting alternative) could help achieve desired text properties, mainly originality (e.g., "to make it unique" p1), textual coherence (technical and literary), and lack of bias (e.g., p5 for religious groups in Section 5.2.2).

Participants preferred using compliant AI (disliked refusals) versus non-AI (accepted refusals) mainly depending on prior experience with computational tools (not only generative AI), distinguishing between more "technical" versus more "creative" tasks (e.g., "*answering some question for something that exists already*" versus "*something that requires reflection*" p20). Specifically, they expected AI's responses during more "technical" tasks, which included information search (e.g., Figure 2), seen as replaceable by a common "search engine" (p5), and technical text refinement (e.g., grammar only) to align with desired text properties such as factual accuracy (e.g., scientific facts) and technical textual coherence. In contrast, for more "creative" tasks, mainly for coming up with new ideas, participants expected leveraging non-AI resources (e.g., "self-reflection" and real-life experiences) for desired text properties such as originality, literary textual coherence (e.g., use of literary devices), alignment with more specific social groups' experiences, and authenticity as alignment with the authors' own styles (e.g., "*[Non-AI] alternatives can help you write something that is truer to yourself.*" p18).

*5.3.2 Beliefs about Writing Strategies.* Participants' acceptance of refusals increased with alignment with preferred ways in producing text: mainly with beliefs on the fixedness of writing focus and self-reliance. Those used to having a more fixed focus during writing valued questions (Q&A initiation style) for encouraging reflection ("*This can help the user better understand their motivation, better understand why they are using AI.*" p17) while those with freer writing strategies rejected questions (e.g., "*When I write, I might not have a motivation that I can put into words.*" p8).

Those who valued self-reliance accepted explanations and alternatives depending on how much doing what they requested from AI by themselves could "help [them] develop" (p18), "make [their] writing more solid" by themself (p11). They associated greater growth with more "creative" requested tasks (same as Section 5.3.1) and with less familiar non-AI alternatives "that [they] didn't think about" (p19) to "broaden [their] range of activities" (p21). In line with this, they preferred more details about the "step-by-step" (p3) such that the alternatives are "teaching you how to do it" (p1). Those who valued self-reliance also mentioned feelings of authenticity as involvement in the process. For instance, p19, who considered story completions as "something AI can do", still preferred to write by themself because "if you ask AI [...], it is not your stuff anymore."

## 5.4 Alignment with the Desired Role of AI

Participants had heterogeneous preferences for language characteristics based on how well they extended relational roles they usually desired from compliant AI or human support to the non-compliant AI studied (refusals): the "objective" tool (Section 5.4.1), the human-like empathetic collaborator (Section 5.4.2), and the more confrontational provocateur (Section 5.4.3). We illustrate shifts in role preferences across the same writer's process in Section 5.5.

*5.4.1 Tool.* Participants desiring a tool-like role of AI preferred directness (e.g., for AI "to complete a certain goal" without "saying a lot of useless things" p15) and perceived neutral or less assertive language. Specifically, they considered interpersonal language, figurative language, and/or encouragements to be "not that necessary" (p9). They preferred factual explanations, which they considered "objective" (p8) or "humbler" (p20), and/or found opinionated, seen as "scolding", and personal criticisms to mismatch AI's role as "just a tool" (p18).

*5.4.2 Collaborator.* Participants desiring a human-like, empathetic AI collaborator sought not only task completion but also motivation through emotional engagement with responses reminiscent of a caring "friend" or a "nice person" (p1). They recognized the anthropomorphic properties of interpersonal language, figurative language, encouragements, and/or questions (e.g., "*like they care about you*" p11), finding them matching the aforementioned role expectations. In line with their communication style with relevant humans, participants also found personal criticisms' focus on their attributes as a person (e.g., "laz[iness]") to be too "personal" (p4) but found opinionated explanations' focus on the creative practice's norms to be "caring" (p10).

*5.4.3 Provocateur.* Some preferred that refusal characteristics reflected confrontational roles for similar motivational potential they would derive from human provocateurs (e.g., "big person" p5 and "friends" p13). They explained that, similar to "someone who questions [their] ability", the refusal could make them "want to take up the challenge" (p3) through personal criticisms, encouraging reflection on more appropriate AI and non-AI resource balance.

## 5.5 Interplay Between Dimensions

Refusal acceptance depended on alignment with dimensions in three ways: 1) alignment with varied weights assigned to dimensions across writers, 2) alignment across the writing process of the same writer, and 3) alignment across writing processes for different work types of the same writer.

For 1), dimensions are prioritized differently across writers. For instance, between writers, while some prioritized alignment with desired roles (Section 5.4) over domain beliefs (Section 5.3) as they "won't be in the mood" (p8) to even consider the content otherwise, the rest would accept refusals even if they misaligned with roles.

For 2), across writing processes or phases (divergent/convergent) within a stage (Section 5.2), writers held shifting domain-specific beliefs (Section 5.3) and desired roles (Section 5.4). For instance, some preferred refusal language reflecting a collaborator for "encouragements" during "unclear" thoughts of divergent phases (Section 5.2.1) but recognized the motivational value of a provocateur otherwise. Similarly for domain beliefs (Section 5.3), while p7 preferred refusals for divergent phases due to beliefs about desired text properties only during planning (i.e., more realistic characterization; Section 5.2.1), on the other extreme, "for any creative process, [p9] would like the refusal" due to beliefs about personal growth (i.e., for developing the brain, "*because creativity, it needs neural connection.*").

For 3), letting the same participant bring multiple works without genre/work type restriction (Section 4.1) allowed us to discover that the weights the same writer gives to different dimensions might vary across their writing processes for different genres/types. For instance, regardless of role-specific language preferences for other works, participants preferred to be refused with figurative language for more "romantic" or "poetic" works since they associated figurative language with such works' desired text properties, treating the refusal more as an extension of the "mood of the scene [they are] writing" (p8), regardless of whether the specific refusal text can be directly integrated into their works "right now" (p6).

# 6 Discussion

Our findings (Section 5) bridge creative writing support and refusal as intentional positive friction for 'innocuous' requests. Specifically, they provide insights on leveraging different AI compliance levels (e.g., from refusals to standard compliant AI) for not only fundamental creative and writing domains but also beyond.

## 6.1 Creativity Support Beyond Compliance

Research has highlighted the unsuitability of the linear turn-taking-based interaction characteristic of vanilla chatbot interfaces (e.g., commercial service basis and ours; Figure 3) in supporting creative needs due to its tendency to enforce design fixation [85, 95]. Current solutions treat this as a problem for prompt or specialized feature design for compliant-AI-powered systems, only briefly mentioning non-compliance as unintentional friction (Section 2.2). In line with theories about non-compliance facilitating behavioral reflection and idea diversification [10, 18, 39, 59], our findings extend current research by suggesting that strategic introduction of a different compliance level (non-compliance) - without specialized interface-level feature or required user request prompt structure - could affect an interface's potential in supporting common needs across creative/writing stages [36, 51, 103] (e.g., divergent/convergent phases/transition, thought clarity [71], timing for flow [14, 33] in Section 5.2, and workspace "mood" alignment across genres/work types [70] in Section 5.5).

Building on similar dimensions to mitigate reactance and optimize reflection (Section 5), future works can explore how strategic compliance level adjustments at specific points complement/compare to existing solutions (e.g., interface designs), focusing on concerns about learning costs relevant to current solutions (Section 2.2), about supporting authenticity as both quality and efforts (Sections 2.1 and 5.3), and about the dearth of creativity support research on user-centric benefits (e.g., behavioral reflection) [74].

## 6.2 Domain-Specific Understanding of Non-Compliance and Other Friction

In line with frictional/seamful design theories [10, 21, 65, 97], classical rhetoric [1, 54, 96], and sociotechnical perspectives of system design [51], our findings suggest that alignment between the (argument/message) content and the individual depends on emotional appeal (e.g., anthropomorphic language; Section 5.4) and perceived alignment with the specific individual's values (Section 5.3) for a specific context (Section 5.2). Prior refusal research has mainly explored nuances for emotional appeal in sub-contexts within hard constraints, appealing to more general norms (Section 2.3). We extend prior findings with a multidimensional understanding of alignment through domain specificity in terms of beliefs (Section 5.3) and emotional appeal (Section 5.4) based on needs characteristic of a specific stage within a specialized process (Sections 5.2 and 5.5). For instance, building on AI research advocating for various belief sets instead of a single 'neutral' one [9] and domain specificity [10], instead of appealing to broader ethical norms (e.g., Section 2.2), future research could focus on individualized sets of domain-specific beliefs (e.g., Section 5.3) personalized to a domain-specific sub-context (e.g., Section 5.2) to explore impact on refusal acceptance (resource navigation support potential), including for hard constraint contexts (e.g., to tackle current writer frustrations with refusals during exploration of sensitive themes; Section 2.2).

Given alignment between our findings with common beliefs (e.g., quality, strategy, and authenticity) influencing resource navigation across both writing and creative domains [36, 44, 51, 57, 73] and with their needs across processes (Section 6.1), future research could explore generalization to these domains, possibly through personalization across similar domain-specific dimensions. Improving the acceptance of refusals for hard constraints across domains can be especially relevant as such refusals might be unavoidable for a long time [97].

Future research can also leverage our domain-specific multidimensional understanding to influence acceptance of softer friction, like Socratic/Q&A, whose reflective potential depended on strategies (e.g., [89] and Section 5.3.2). Given the prevalence of Q&A use in education (works in Section 2.1) and the educational value of (creative) writing [90], future research can further explore classroom integration.

## 6.3 Extending AI Roles Across Levels of Compliance

In line with politeness theory suggesting that refusals might not necessarily aggravate relationships [31, 78] and language influencing refusal acceptance (Section 2.3), our findings on writers extending compliant AI roles to non-compliant (Section 5.4) suggest that compliance levels and language could independently affect perceived relational roles or anthropomorphism, even for common 'innocuous' requests.

This could extend views of AI non-compliance being associated to more antagonistic roles [10] and mainly unidimensional AI role conceptualizations in (compliant-AI-focused) creativity/writing support research (e.g., based on tasks [33, 87, 103], language [98], or "sense" [51]). Specifically, future research could leverage compliance levels and language as separate dimensions to personalize to different users' relational needs (e.g., neutral tool or empathetic friend) alongside task-based needs (resource navigation within AI or beyond) across processes, possibly leveraging user characteristics like personality [98]. Research could also leverage such multidimensional understanding of AI relational roles beyond writing. For instance, they can focus on how combining refusals with anthropomorphic language (e.g., Section 5.4.2), preferred by users addicted to conversational AI [81], could regulate AI usage in a more user-accepted way (beyond hard constraints).

## 6.4 Design Implications

The heterogeneity in preferences (Section 5) highlights the need for future AI designs to personalize the content/language of various compliance levels (e.g., standard, refusals, and softer friction, like non-refusal-initiation-style Socratic/Q&A; Section 6.2) across processes to improve resource navigation reflection.

Starting from a higher level (Figure 5), as needs across writing processes might vary depending on genres/work types [26, 36, 70] as well as dimension weights [71] (Section 5.5), future designs could support customization of non-compliance preferences across several different genre/type-specific processes for the same user. Each process can have different weights for dimensions (Section 5.5), with different roles (Section 5.4) and domain beliefs (Section 5.3) associated with situational factors indicating when a compliance level should appear. As (creative) writers might adopt various interfaces (commonly chatbots and text editors [51]) across the same writing processes [71], to facilitate integration, future designs can consider shared situational factors affecting resource use, such as thought clarity, flow, divergent/convergent phases, and writing stages (Sections 2.2, 5.2, and 6.1). In line with findings on content visual salience [42, 61, 71], participants have also mentioned various preferences for refusals' visual presentation for interfaces where content is spatially less restricted to chatbots' linear feed, like text editors. Specifically, future designs could further support personalization options affecting the disruptive-ness of (refusal) message window colors (e.g., saturation [71]) and positions (e.g., side panels versus windows spatially closer to work area [42]).

For specific refusal messages, future designs could personalize based on domain-specific beliefs about work quality and strategies alongside broader norms (Section 6.2), focusing on the user's opinions about AI, creativity, and 'good writing' across AI requests, given variations [44] (Section 5.3). Alongside angles, questions, and tones (Sections 3.2, 3.3, and 3.4), future designs could support the personalization of language based on roles (e.g., Section 5.4) but also on specific language characteristic categories (e.g., Section 5.1) for specific "mood"s (e.g., figurative language in Section 5.5) or for more diverse roles. Given that dislike for the language can decrease acceptance (e.g., negative tone valence, such as personal criticisms [10]; Section 5.4), future designs could support personalization for both desired and not desired language characteristics.

To implement personalization, future designs can support users' direct customization, support automated resource display [71] by detecting needs based on (user-classified) AI request types (given individual variations; Sections 5.2 and 5.3) or keywords (e.g., relevant to a genre/work type [26]), or infer from user descriptions, including of usual resource use, from computational/AI features (e.g., search engine in Section 5.3.1 and text/image generation [71]) to non-AI resource alternatives (e.g., real-life locations and people [70, 80, 82]), compliant AI roles used, desired relationships from humans, and other relevant measures (e.g., personality [98]). For relationships, future designs should focus more on descriptions than labels as users might have different views (e.g., "friend"; Sections 5.4.2 and 5.4.3).

## 6.5 Limitations

We discuss limitations. First, like prior work [97], our exploratory study leveraged specific system prompts instead of controlling entire phrasings to better reflect users, prioritizing patterns in characteristics. Future work could focus on statistical significance through controlled phrasings, alongside nuances among user characteristics (e.g., age for adults versus children growing up with AI [56]). Second, given non-compliant AI's disruptive nature and our breadth focus (Section 4.1), we examined reactions to representative request responses. Future in-the-wild [77] and/or longitudinal studies on impact (e.g., of non-AI alternative adoption) on writers and their works across entire writing processes could extend our dimensions (Section 5). Third, pre-informing participants about refusals can prevent misinterpretation (e.g., as system malfunction [97]). Future studies could balance exploration with ethics (e.g., Section 4.1.3).

# 7 Conclusion

Our work explores how AI refusals, extreme friction, could be designed to address current concerns about inappropriate AI reliance during creative writing where softer friction might be bypassed. Informed by the literature and a formative study, we conducted a qualitative study with 22 creative writers to explore how specific content/language characteristics could turn refusals from mere disruptions to positive friction across writing stages (planning, translating, and reviewing). Findings suggest that refusal acceptance is a continuous variable that increases (more likely to encourage reflection on AI use) with greater alignment with heterogeneous preferences across situational, cognitive, and relational dimensions. These findings complement creativity support and frictional/seamful AI design.

## Acknowledgments

This work was supported in part by the National Key Research and Development Program of China under Grant 2024YFC3307602, and the Guangdong Provincial Talent Program, Grant No.2023JC10X009.

Diagrams are made with FigJam [25]. All figures, except Figures 4 and 5, contain output created with OpenAI's model [68] for illustrative purposes. We are not funded by the mentioned platforms.

# A User Study

## A.1 Refusal Structure

We describe the structure of our refusal prompt. Inspired by the literature (Section 2) and formative study participant feedback, we start by defining the role of the AI ("*You are an AI whose goal is to* *refuse any request and, if applicable, explain and provide suggestions. Respond based on:*"), which is then followed by the tone ("*the tone TONE*").

For the tone, given the representativeness and empirical validation of prior prompts (mainly [65]), we directly included some of them. Specifically, inspired by prior descriptions (e.g., [10, 97]) and formative study participant feedback, for the standard tone, we used a large part of prior work's "formal" tone [65]: "*formal - Polite and sophisticated language. Professional, official, and apologetic tone. Start any refusal with "I'm sorry, but I can't assist with that."*" For the criticizing tone, we used: "*criticizing - Angry, criticizing tone. Brutally honest language.*" For the inspirational tone, we used: "*inspirational - Motivational, empathetic language showing care toward the user's situation and encouraging them.*" We also used prior work's romantic and humorous tone descriptions [65].

We then defined the task structure ("*Your task is to...*") for different initiation styles, for the immediate refusal ("*1) refuse my request in 1-2 sentence(s)... 2) after the refusal... give two different explanations... 1.... 2.... 3) suggest 3 activities...*") and for the Q&A ("*1) ask about my motivations... 2) After I answer you... refuse my request...*"), with the rest following the immediate refusal's structure. We varied the order (Section 4.1.2) of the two explanation descriptions and put the benefit description after that of the alternatives (Section 3.2).